\documentclass[AMA,STIX2COL,intlimits,sumlimits,namelimits,centertags]{MRM}
\articletype{Research Article}%

\received{25 January 2024}
\revised{TBD}
\accepted{TBD}
\usepackage{cancel}

\usepackage{mathdots}
\begin{document}

\title{Improved Image Reconstruction and Diffusion Parameter Estimation Using a Temporal Convolutional Network Model of Gradient Trajectory Errors}

\author[1]{Jonathan B. Martin}{\orcid{0000-0002-9384-8056}}

\author[1,2]{Hannah E. Alderson}{}

\author[1,2,3]{John C. Gore}{}

\author[1,2,3]{Mark D. Does}{}


\author[1,2]{Kevin D. Harkins}{}

\authormark{JONATHAN B. MARTIN \textsc{et al}}

\address[1]{\orgdiv{Vanderbilt University Institute of Imaging Science}, \orgname{Vanderbilt University Medical Center}, \orgaddress{\state{Tennessee}, \country{USA}}}

\address[2]{\orgdiv{Department of Biomedical Engineering}, \orgname{Vanderbilt University}, \orgaddress{\state{Tennessee}, \country{USA}}}

\address[3]{\orgdiv{Department of Radiology and Radiological Sciences}, \orgname{Vanderbilt University Medical Center}, \orgaddress{\state{Tennessee}, \country{USA}}}

\corres{Jonathan B. Martin \email{jonathan.bach.martin@vumc.org}}

\presentaddress{1161 21st Ave. S., AA-1105, Nashville, TN, 37232}

\finfo{This work was supported by \fundingAgency{National Institute of Health} grants \fundingNumber{T32 EB001628-21} and \fundingNumber{R01 EB031954}}
 
\abstract[Summary]{Errors in gradient trajectories introduce significant artifacts and distortions in magnetic resonance images, particularly in non-Cartesian imaging sequences, where imperfect gradient waveforms can greatly reduce image quality.
\section{Purpose}  Our objective is to develop a general, nonlinear gradient system model  that can accurately predict gradient distortions using convolutional networks.
\section{Methods} A set of training gradient waveforms were measured on a small animal imaging system, and used to train a temporal convolutional network to predict the gradient waveforms produced by the imaging system. 
\section{Results} The trained network was able to accurately predict nonlinear distortions produced by the gradient system. Network prediction of gradient waveforms was incorporated into the image reconstruction pipeline and provided improvements in image quality and diffusion parameter mapping compared to both the nominal gradient waveform and the gradient impulse response function. 
\section{Conclusion} Temporal convolutional networks can more accurately model gradient system behavior than existing linear methods and may be used to retrospectively correct gradient errors.
}

\keywords{image reconstruction, diffusion MRI, machine learning, temporal convolutional network, noncartesian trajectory, time series forecasting, gradient correction}

\wordcount{XXX}

\jnlcitation{\cname{%
\author{J.B. Martin}, 
\author{H. E. Alderson}, 
\author{J.C. Gore}, 
\author{M. D. Does}, and 
\author{K.D. Harkins}} (\cyear{2024}), 
\ctitle{Improved Image Reconstruction and Diffusion Parameter Estimation Using a Temporal Convolutional Network Model of Gradient Trajectory Errors}, \cjournal{Magn. Reson. Med.}, \cvol{2017;00:1--6}.}

\maketitle
\textcolor{red}{\textbf{\huge Submitted to Magnetic
Resonance in Medicine}}


\section{Introduction}\label{intro}

Magnetic resonance imaging (MRI) pulse sequences frequently incorporate time-varying gradient waveforms that put great demands on associated gradient hardware. Examples include gradient waveforms for non-Cartesian imaging \cite{Willmering2020ImprovedMRI, Shen2023Ultra-shortPattern}, selective excitation localization \cite{Tong2020ImprovingStrategy, Luo2021JointMRI}, or diffusion encoding \cite{Xu2021ProbingHumans, Chakwizira2023DiffusionExchange}. These applications require high fidelity of the gradient system, as distortions of the gradient waveform  reduce image quality, for example, causing artifacts after reconstruction \cite{Mani2018AImaging} or distortion of the excitation profile \cite{Takahashi1995CompensationTrajectories}. 

Gradient distortions may be linear or non-linear in nature. For example, system group delays\cite{Bhavsar2014FastMRI} and system bandwidth limitations result in waveform distortions that are time-invariant and linear in both spatial and time dimensions. Eddy current field distortions are temporally linear, but may be spatially nonlinear \cite{Jezzard1998CharacterizationImaging, Li2011FiniteMagnet}. Other distortions are known to be more challenging to correct, and may be the product of system characteristics which are not linear-time-invariant (LTI) . Additionally, gradient amplifiers themselves are nonlinear devices. Gradient amplitude, slew rate, and acceleration constraints imposed by the hardware are nonlinear effects. Other nonlinear amplifier distortions can occur within the normal operating range, such as amplifier droop \cite{Babaloo2022nonlinearSystems} or zero-crossing distortions in two-stage push-pull amplifiers such as the H-bridge, which are popular in gradient amplifier design \cite{Sabate2005High-bandwidthControl, Zeng2022High-PrecisionAmplifiers}. Further, gradient hardware heating results in a time-varying gradient system response \cite{Stich2020TheCharacteristics, Nussbaum2022ThermalModeling} and vibrations due to mechanical resonances can produce nonlinear field distortions \cite{Zhang2014VibrationalMagnets}. 

One strategy to improve the fidelity of the gradient system is to directly measure the gradient waveforms---either through pulse sequence methods \cite{Onodera1987AImaging, Harkins2021EfficientVariable-prephasing} or with monitoring hardware \cite{Barmet2008SpatiotemporalMR, Vannesjo2013GradientCamera}---and then use the measured gradient trajectories to improve image reconstruction and/or RF pulse design. While this approach is accurate, it can be time-consuming, requiring calibration scans to be acquired when there is any modification to the gradient waveforms. Furthermore, gradient monitoring hardware incurs an additional system expense.

An alternative approach is to create a system model of the gradient hardware chain, which may then be used for retrospective correction\cite{Vannesjo2016ImagePrediction, Lee2022MaxGIRF:Effects}, or to prospectively design pre-emphasized gradient waveforms that produce nominal temporal gradient waveforms after they are corrupted by the system \cite{Ahn1991AnalysisImaging}. Most modern MRI systems provide methods to prospectively compensate for group delay and eddy currents \cite{Peters2003CenteringErrors, Robison2010FastMRI, Bhavsar2014FastMRI,Jensen1987ReductionInstrument, Morich1988ExactSystems}. 

More recently, a linear-time invariant model such as the gradient impulse response function (GIRF) or the equivalent frequency domain gradient system transfer function (GSTF) has been proposed\cite{Vannesjo2013GradientCamera, Addy2012AEstimation, Campbell-Washburn2015Real-timeFunction} to model the gradient system. However, these linear models are incapable of accurately correcting the behavior of a gradient system which is not linear time-invariant (LTI). In some systems, nonlinear behavior may be particularly pronounced and severely impact image quality despite GIRF correction \cite{Rahmer2021Non-CartesianCurrents, Scholten2024OnInsert, Babaloo2022nonlinearSystems}. Attempts have been made to expand the GIRF model to account for nonlinear temperature effects \cite{Nussbaum2022ThermalModeling}, but this model is still an adaptation of the original linear model and is incapable of properly characterizing nonlinear behavior. 
\par In contrast, deep learning networks provide an opportunity to predict nonlinear artifacts of the gradient system. Recurrent neural networks such as the long short-term memory (LSTM) architecture have previously been used \cite{Hochreiter1997LongMemory, Alazab2020AGrid, Chen2021ResearchNetwork, Liu2022GradientNetwork} for sequence modeling. Convolutional neural network (CNN) architectures have been shown to have some advantages over recurrent networks in this area in terms of reduced memory and time costs, easier parallelization in training, and improved stability and generalization in many cases \cite{Bai2018AnModeling, Pascanu2013OnNetworks, Li2025APrediction}. In particular, temporal convolutional networks (TCN) \cite{Lea2017TemporalDetection} have had success in modeling complex physical systems \cite{Yan2020TemporalENSO, Lin2021TemporalForecasting, Li2025APrediction}. Despite the popularity of deep learning approaches for time series forecasting and prediction in nonlinear systems, there has  been only preliminary work on deep learning-based modeling of MRI gradient systems \cite{Liu2022GradientNetwork}. Hence, the application of deep learning to this problem and the demonstration of real-world applicability of deep learning models has yet to be fully assessed. 

\begin{figure*}
\centerline{\includegraphics[width=18cm,keepaspectratio]{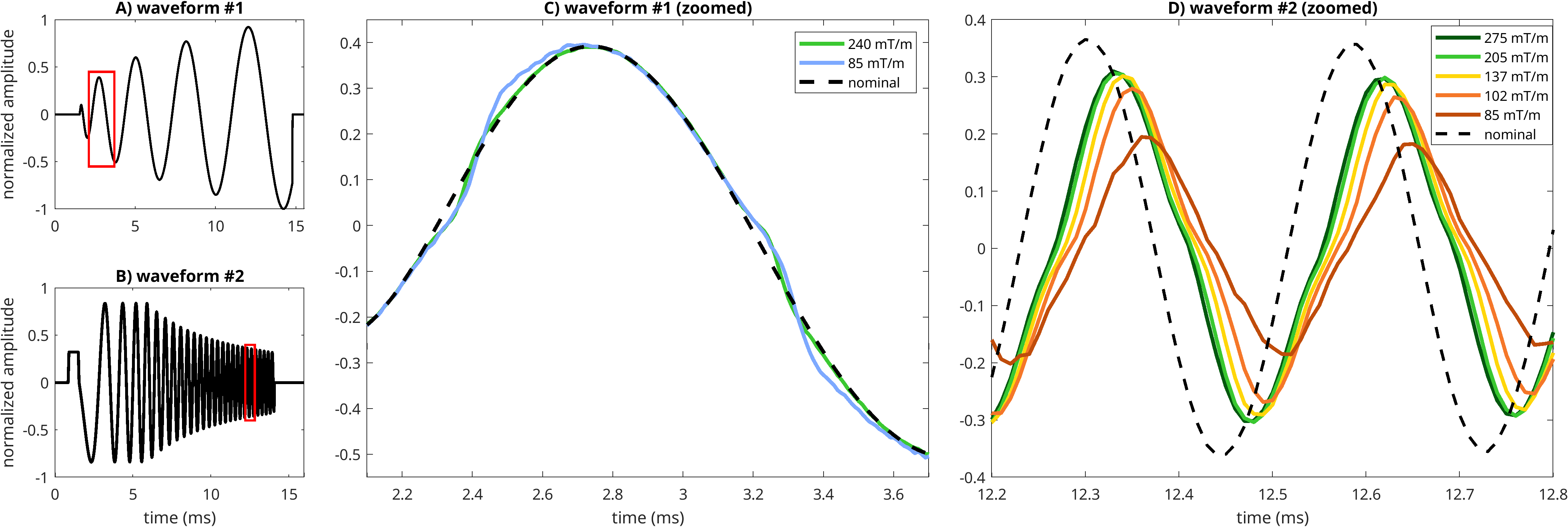}}
\caption{Spiral and chirp gradient waveforms measured on the 7T Bruker system, with conspicuous nonlinearities. Waveform timecourses are normalized by their respective maximum nominal amplitudes. A) and B) show the nominal waveforms with zoomed-in ROI highlighted. C) and D) show the nominal and measured waveforms at several different amplitudes. Clear nonlinearities are present. In C) the two waveforms are not simply scaled copies of one another but have distinct zero crossing artifacts. In D), the response varies with amplitude of the applied waveform. Delay and attenuation increase with decreasing amplitude.  \label{fig2}}
\end{figure*}

To overcome the shortcomings of LTI methods for the modeling of gradient systems, we propose a learned model of the full gradient system to predict readout gradient waveforms. The model is created using pulse-sequence-based measurements of a training library of gradient waveforms. We show that a learned model of the gradient system can accurately predict gradient waveform shapes not seen in the training process, demonstrating the feasibility of applying learned models to real-world MRI systems. The utility of the model for improving the quality of the reconstructed image and the estimation of quantitative parameters is demonstrated on a preclinical 7 Tesla system.

\section{Methods}
We examined the feasibility of using a neural network to predict gradient waveform errors and the effectiveness relative to established linear methods. First, a library of gradient waveforms was measured to characterize the gradient system response. These measurements were used to construct both a linear (GIRF) and neural network-based (TCN) model of the gradient system. Both models were used to predict readout gradient errors, and the predicted gradient waveforms were used in the reconstruction of multi-shot spin-echo images and diffusion parameter maps. 

All MR experiments were performed on a 7T 16 cm bore Bruker Biospin Avance III HD small animal scanner (Bruker BioSpin GmbH \& Co. KG, Ettlingen, Germany). A 38mm transmit/receive quadrature volume coil was used for RF excitation and reception. The scanner is equipped with actively shielded gradients with a maximum gradient strength of 770 mT/m and rise times of 146 \textmu{}s to full amplitude. Built-in eddy current compensation was enabled for all experiments.

\subsection{Training and Validation Gradient Waveform Collection}
A library of 18 gradient waveforms was designed to characterize the response of the gradient system and provide training and validation data for the network. Data files with all training waveforms are stored at the following link: \url{https://github.com/jonbmartin/tcn_gradient_prediction}. The training and validation gradient waveforms contained realistic readout waveforms for imaging, such as spiral and trapezoidal waveforms, and additional waveforms designed to characterize the response of the system, such as chirps, multisines, and triangle waveforms\cite{Vannesjo2013GradientCamera}. 11 waveforms were assigned to the training dataset, and 7 were withheld for validation. The training set primarily consisted of system characterization (non-imaging) waveforms, while the validation set included primarily imaging waveforms. The readout waveforms used in the imaging experiments comprised a third distinct test dataset, non-overlapping with the validation set. 

All gradient waveforms were measured on all 3 gradient axes using a previously established pulse-sequence based method\cite{Harkins2021EfficientVariable-prephasing}. Measurements were performed using a homogeneous 5 mL spherical phantom filled with a 5mM CuSO\textsubscript{4} solution. The variable pre-phasing sequence had TE/TR = 10.8/500 ms and used 8 variable pre-phasing steps across 9 slices with a thickness of 0.35 mm. 15 gradient waveform amplitudes uniformly distributed in the range [70, 700] mT/m were measured for waveforms in the training dataset, which resulted in a scan time of 8 minutes/waveform. Only 7 gradient waveform amplitudes in the same range were measured for validation waveforms, which resulted in a scan time of 3m 44s/waveform. Final training/validation split of the data based on the total number of measured time points of data in each dataset was 76\% train, 24\% validation. The total scan time to measure all amplitudes for 11 unique training waveforms and 7 unique validation waveforms on all 3 gradient axes was 5 h 42m. Gradient waveforms from the training dataset were used to form a GIRF model\cite{Vannesjo2013GradientCamera} of the gradient system for all 3 axes. Unlike in the original GIRF method\cite{Vannesjo2013GradientCamera} gradient waveforms were measured using the a pulse-sequence based method\cite{Harkins2021EfficientVariable-prephasing} rather than with a dynamic field camera.  

\par Examples of measured gradient  validation waveforms are given in Figure \ref{fig2}, including a spiral readout waveform (A) with zero-crossing nonlinearities that are amplitude dependent (C). Similarly, a chirp waveform is also shown (B), exhibiting amplitude dependent variations in the magnitude and group delay of the measured response waveform (D). 

\subsection{Network Architecture and Training}
\begin{figure*}
\centerline{\includegraphics[width=16cm,keepaspectratio]{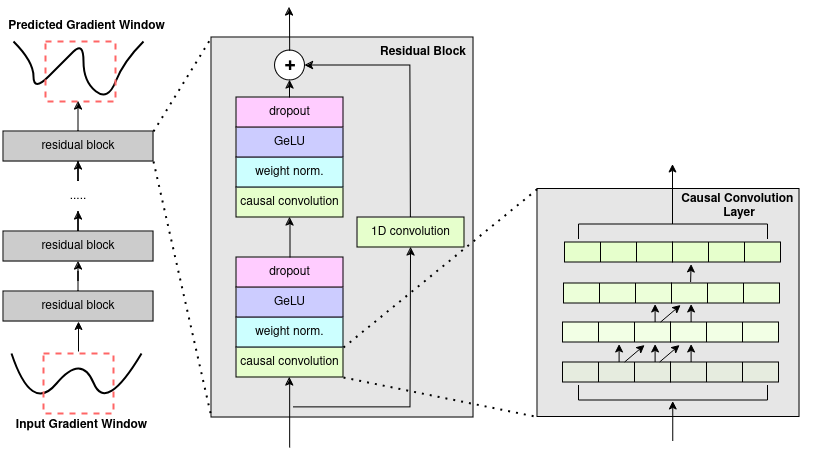}}
\caption{The temporal convolutional network architecture\cite{Bai2018AnModeling} used in this work. The TCN consists of multiple concatenated residual blocks. Each residual block consists of two layers of causal convolution, each followed by weight normalization, GeLU activation, and a dropout layer. A 1D residual convolution bypasses the causal convolution layers. The causal convolution prevents violations in causality of the data by forcing convolution outputs to only have dependencies on prior timepoints. \label{fig1}}
\end{figure*}

The TCN architecture used for gradient waveform prediction is shown in Fig. \ref{fig1}, and follows the architecture of a multistage TCN outlined in Ref. [\citen{Bai2018AnModeling}]. This TCN takes in a window of size $[N_t, 2]$, containing $N_t$ samples of the normalized [-1, 1] nominal gradient waveform and the system's scaling of the gradient waveform also normalized to [-1, 1]  at each time point (typically constant for a given gradient pulse). This formatting was chosen to prevent dynamic range issues due to large differences in waveform scaling within the dataset, either in input value or target value \cite{LeCun2012EfficientBackProp}.  The fundamental building block of the TCN is the causal convolution layer--- a 1D convolution layer of size $T$ which uses a kernel of size $k$ incorporating only previous time points in the calculation of output values in order to preserve the temporal order of data. To capture long-time dependencies, a dilated convolution is used, which allows the receptive field of the convolution to grow exponentially without a corresponding increase in model size or computational cost. Two causal convolutions are concatenated within residual blocks, with weight normalization, a GELU activation layer\cite{Hendrycks2016GaussianGELUs}, and a dropout layer following each convolution. Residual blocks also incorporate a 1D convolution with a small $k$=1 kernel that bypasses the causal convolutions. The full TCN consists of $N_{res}$ cascaded copies of these residual blocks. A final linear layer returns the normalized gradient sample prediction at time point $t$. The network was trained using a mean absolute error loss of the prediction values. 

To find optimal network hyperparameters within this architecture, a hyperparameter optimization was performed using the Optuna toolbox\cite{Akiba2019Optuna:Framework} with the measured train and test waveforms. The default Tree-structured Parzen Estimator algorithm was used to find optimal hyperparameters. The final network hyperparameters were $N_{res}$ = 5, $k$ = 16, dropout probability= 0.2, and $T$ = 48. We used an exponential dilation for the convolutional layers of $d = 2^i$ for the ith residual block. Optimizer hyperparameters for an Adam optimizer were a learning rate of $8.4\times10^{-4}$, a batch size of 32, a weight decay of $9\times10^{-8}$, $\beta_1= 0.9$, $\beta_2=1-1\times10^{-6}$. All other parameters used were PyTorch defaults. Using these hyperparameters, the network was trained for 150 epochs, with a unique model trained for each gradient axis. 

\par A long short-term memory (LSTM) recurrent network \cite{Hochreiter1997LongMemory, Alazab2020AGrid, Chen2021ResearchNetwork, Liu2022GradientNetwork} was also examined to assess the viability of alternative network architectures that are used for time series forecasting tasks. The Optuna toolbox was again used to perform a hyperparameter optimization, which produced the following optimal network: 2 layers with hidden layer size 96, learning rate of $8.5\times10^{-4}$, a batch size of 32, a weight decay of $2\times10^{-8}$, $\beta_1= 0.8$, $\beta_2=0.99$. All other parameters used were PyTorch defaults. Using these hyperparameters, this network was also trained for 150 epochs. As discussed in later sections, comperable but slightly inferior performance was observed with the LSTM network as compared with the TCN, so it was not used in further studies.

\par Training and validation were performed on a Nvidia RTX A5000 GPU (Nvidia Corporation, Santa Clara, CA, USA) with 24 GiB memory. Training each gradient axis model took approximately 2 hours. Test waveform prediction (multishot spiral and rosette) was performed on a local System76 (System76, Inc., Denver, CO, USA) Adder CPU workstation with 5.4 GHz Intel i9-13900HX and 64 GB RAM. Prediction time for all 80 spiral in-out gradient waveforms (40 shots with 2 waveforms per shot) was 18.27 s or 0.23 s/waveform. Prediction time for the rosette readouts (also 80 waveforms) was 22.02 s or 0.28 s/waveform related work, we have made network code and a library of training, validation, and test gradient waveforms available at \url{https://github.com/jonbmartin/tcn_gradient_prediction}. A table summarizing the waveforms used can be found in the GitHub repository or in the online supplemental materials as Table S1.

\subsection{Imaging Experiments} 

Images were acquired of a phantom and an \textit{ex-vivo} ferret brain. The phantom consisted of a plastic toy brick embedded in a mixture of $CuSO_4 \times 2H_2O$ salt solution (1 g/L) and agar gel (10 g/L) (Bruker standard phantom: T10681). After euthanasia in compliance with a protocol approved by the Vanderbilt University Insitutional Animal Care and Use Committee, the ferret brain was first perfused  with 1L of chilled phosphate buffered saline (PBS, 1X) with 47.6 mg of heparin and then with 1L 2\% PFA, 2.5\% glutaraldehyde in 1X PBS with 47.6 mg of heparin and 0.25 mM of gadolinium (ProHance). Tissues were post fixed and then stored long term in 1X PBS, 1 mM gadolinium and 0.01\% sodium azide prior to imaging.

Two different multi-shot noncartesian readout trajectories were used for imaging. The first was a 40-shot spiral-in-out trajectory (with a readout duration of 5.27 ms), and the second was a 40-shot 5-petal rosette trajectory (with a readout duration of 6.71 ms) \cite{Noll1997MultishotImaging}. 
These trajectories were chosen for this study because it was anticipated that they would be highly sensitive to gradient distortions. This is due to k-space DC measurement occurring at a late timepoint (or timepoints) in both trajectories, which allows for gradient errors to accumulate resulting in inaccurate measurement of high-signal regions of k-space.
These readout trajectories were incorporated into a multi-shot spin echo pulse sequence (TR/TE=1000/10 ms, FOV=30$\times$30 mm, 0.3$\times$0.3$\times$0.5 mm resolution). In the ferret brain, a multi-shot spin-echo DTI sequence (TR/TE=1000/20 ms, FOV=30$\times$30 mm, 6 diffusion encoding directions, b = 1000 s/mm$^2$) was also acquired. Imaging was repeated without diffusion encoding gradients. In all cases, 15 axial slices were acquired.

\subsection{Image Reconstruction and Diffusion Analysis}

Reconstruction was repeated using nominal, GIRF-predicted, TCN-predicted, and measured trajectories. Raw signal was density compensated \cite{Pipe1999SamplingSolution} and gridded based upon the estimated k-space trajectories. $B_0$ maps were derived from the multi-shot rosette spin echo pulse sequence data \cite{Liu2021Myocardial1.5T}, and incorporated into reconstruction using a time-segmented off-resonance correction\cite{Noll1991AGradients}. 

DTI parameters were estimated with weighted linear least squares fitting \cite{Kingsley2006IntroductionOptimization} using the REMMI-Matlab toolbox (\url{https://remmi-toolbox.github.io/})\cite{Harkins2017REMMI-Matlab}. The mean diffusivity (MD), fractional anisotropy (FA), radial diffusivity (RD) and axial diffusivity (AD) were estimated for images acquired with each trajectory. 

\section{Results}

\begin{figure*}
\centerline{\includegraphics[width=16cm, keepaspectratio]{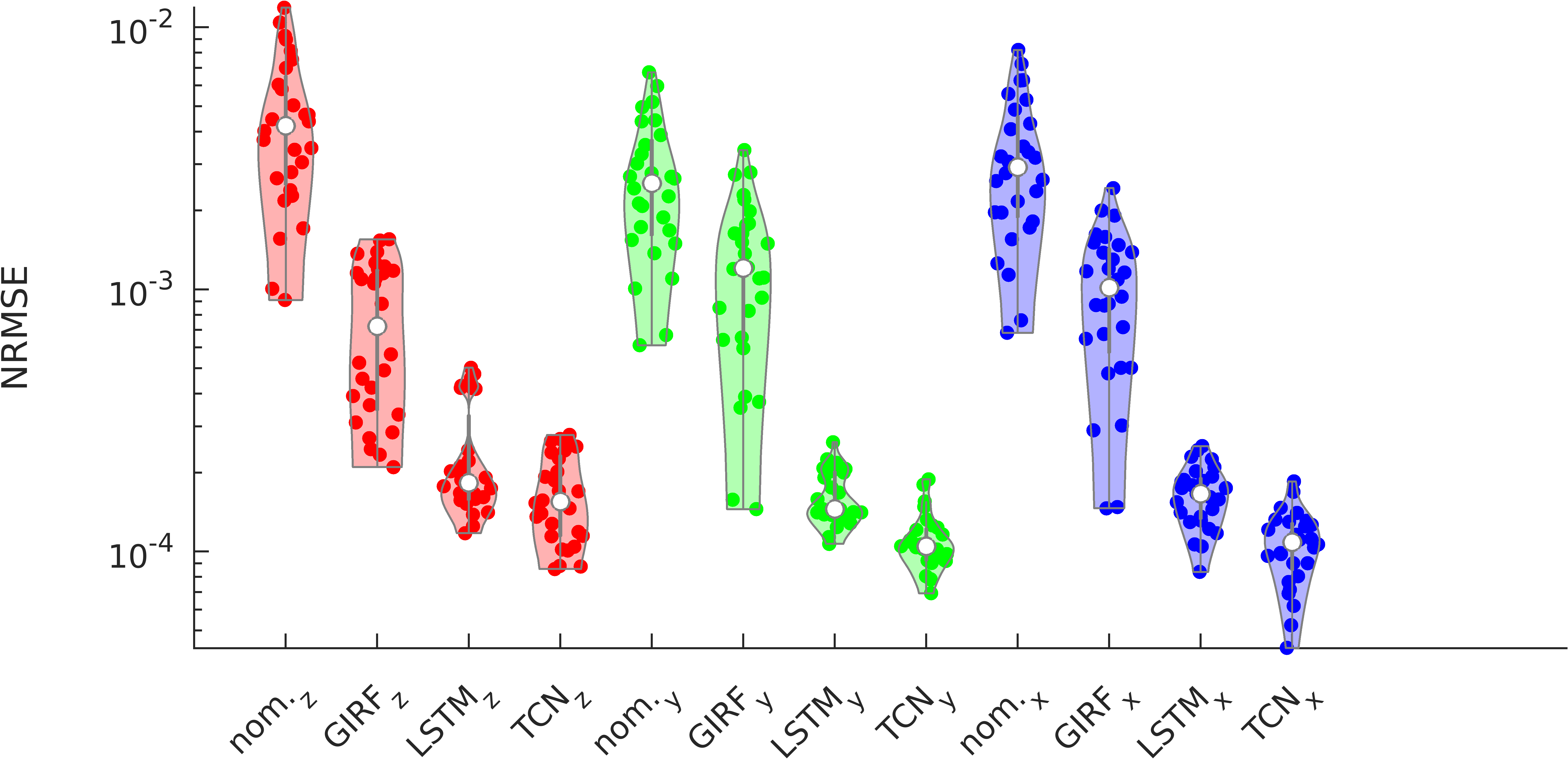}}
\caption{Violin plot of NRMSE for individual validation set waveforms, for the nominal waveform and for each model (GIRF, LSTM, and TCN) on each axis (z, y, x). The GIRF reduces waveform error by approximately one order of magnitude, with either LSTM or TCN modeling reducing error by approximately one additional order of magnitude.\label{fig_violin}}
\end{figure*}

Figure \ref{fig_violin} shows the NRMSE between the predicted waveforms and the measured waveforms for individual pulses in the validation set. Models created for all 3 gradient axes are shown. GIRF estimates of gradient waveform distortions reduced time-domain waveform error on all three axes, while using either TCN or LSTM models further reduced time-domain waveform error. Mean nominal waveform NRMSE for $G_z$ = $4.19\times10^{-3}$, $G_y$ = $2.53\times10^{-3}$, and $G_x$ = $2.92\times10^{-3}$. Mean GIRF-predicted waveform NRMSE for $G_z$ = $7.21\times10^{-4}$, $G_y$ =$1.20\times10^{-3}$, and $G_x$ = $1.01\times10^{-3}$. Mean LSTM-predicted waveform NRMSE for $G_z$ = $1.82\times10^{-4}$, $G_y$ = $1.45\times10^{-4}$, and $G_x$ = $1.66\times10^{-4}$. Mean TCN-predicted waveform NRMSE for $G_z$ = $1.55\times10^{-4}$, $G_y$ = $1.04\times10^{-4}$, and $G_x$ = $1.08\times10^{-4}$. Overall, the TCN gradient system model produced the lowest error for waveforms on all three axes.

\par
\begin{figure*}
\centerline{\includegraphics[width=16cm, keepaspectratio]{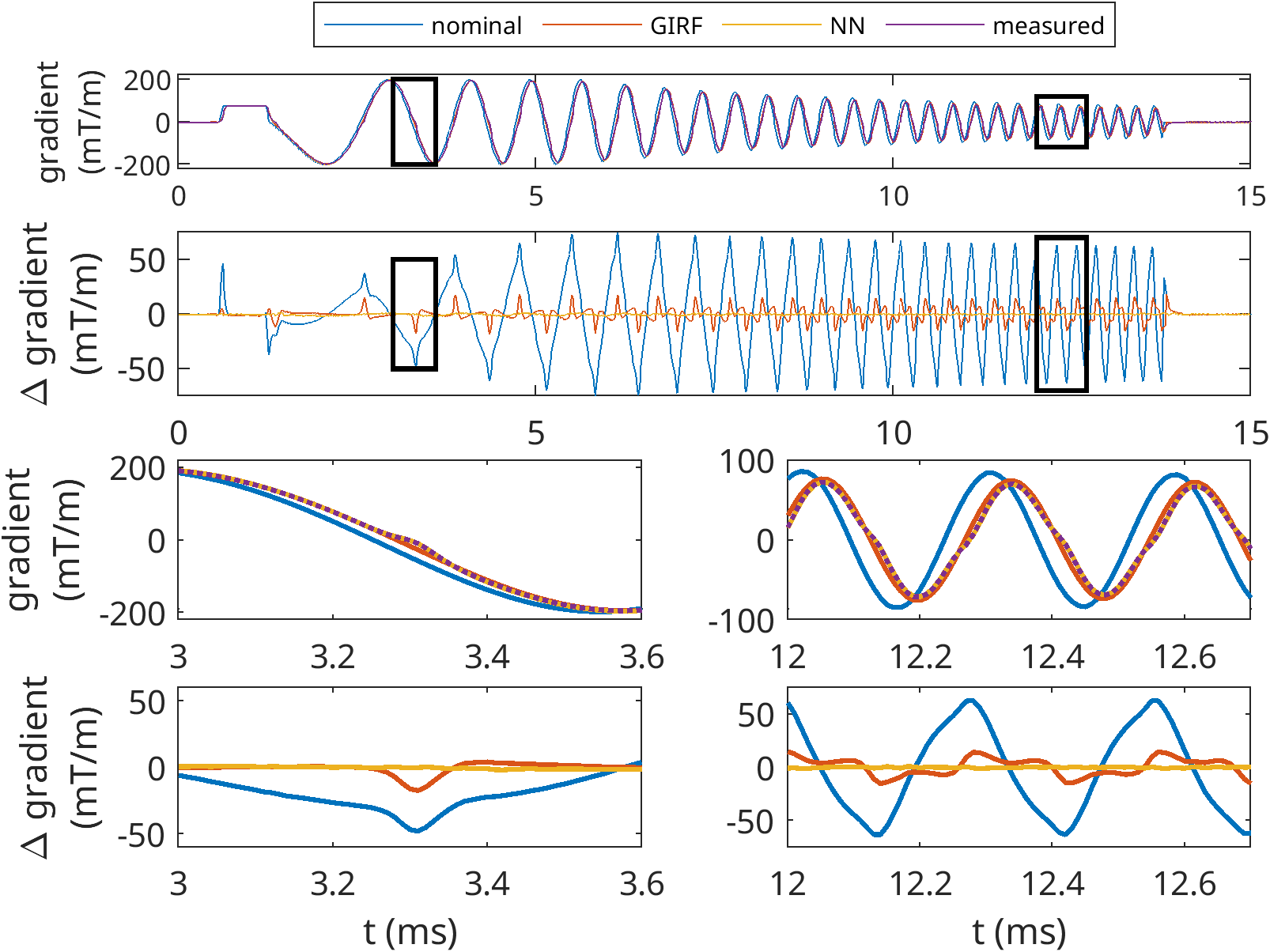}}
\caption{Gradient waveforms for each prediction method, on validation-set chirp waveform. Top: Full nominal, GIRF-predicted, TCN-predicted, and measured gradient waveforms, and errors ($\Delta$). Boxes indicate ROIs below. Bottom left: Zoomed in 0.25 ms time segment of waveform and errors, exhibiting zero-crossing nonlinearity. Bottom right:  zoomed in 0.25 ms time segment of waveform and errors, exhibiting errors that the GIRF does not accurately predict. \label{fig3}}
\end{figure*}

 Figure \ref{fig3} highlights time-domain errors between the measured and predicted waveform in the same chirp waveform as Figure \ref{fig2}. The GIRF reduced waveform error, but struggled to accurately characterize the nonlinearities in the system response --- including both the zero-crossing distortions shown in the bottom left panels of Figure \ref{fig3} and the changing system response with input waveform frequency, which manifested primarily as a reduction in measured waveform amplitude relative to input waveform at higher frequencies, shown in the bottom right panels of Figure \ref{fig3}. The TCN model is better able to account for both of these phenomena. RMSEs of the nominal, GIRF-predicted, and TCN-predicted trajectories were 34.43, 6.08, and 0.63 mT/m respectively. 

\begin{figure}
\centerline{\includegraphics[width=11cm, keepaspectratio]{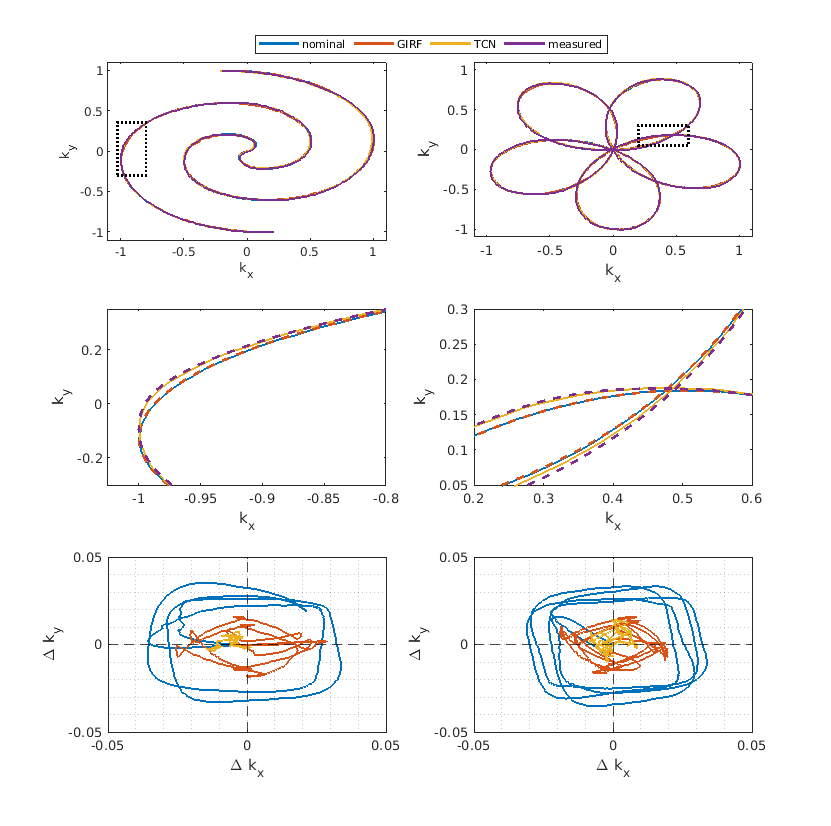}}
\caption{Measured k-space trajectories and trajectory errors for one of the 40 shots of the rosette and spiral trajectory readouts, normalized to the maximum k-space coordinate. A-B) show full trajectories, C-D) zoomed in regions on each trajectory, E-F) 2D k-space errors for nominal, GIRF-predicted, and TCN-predicted trajectories. \label{fig4}}
\end{figure}

Figure \ref{fig4} shows the k-space trajectory estimates and errors for a representative single shot of the spiral and rosette readout waveforms used in the imaging experiments. The GIRF and TCN models both predicted a trajectory that was more accurate to the measured values than the nominal trajectory, but the TCN had a greater error reduction than the GIRF. Mean RMSE values in normalized k-space units across all 40 spiral shots were $(2.23\pm0.12)\times10^{-2}$, $(1.38\pm0.64)\times10^{-2}$, and $(0.82\pm0.12)\times10^{-2}$ for the nominal, GIRF-predicted, and TCN-predicted trajectories. Mean RMSE values across all 40 rosette shots were $(1.96\pm0.13)\times10^{-2}$, $(1.10\pm0.08)\times10^{-2}$, and $(0.45\pm0.21)\times10^{-2}$ for the nominal, GIRF-predicted, and TCN-predicted trajectories.

\begin{figure}
\centerline{\includegraphics[width=8.5cm, keepaspectratio]{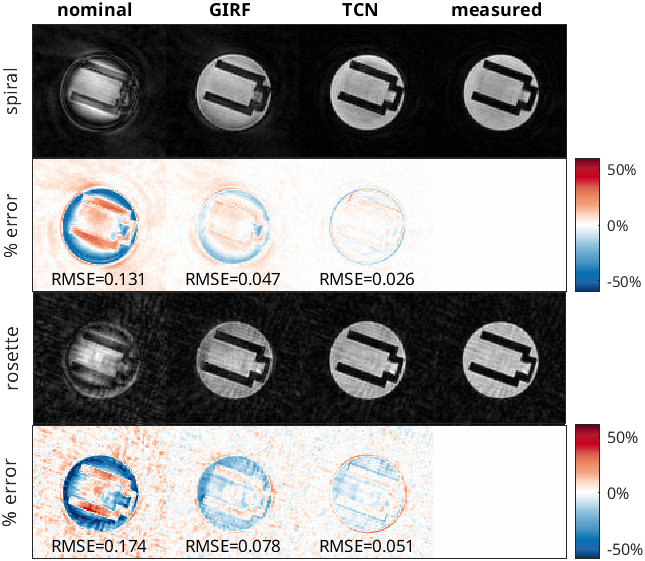}}
\caption{Phantom images acquired with spiral and rosette readout waveform, reconstructed from nominal, GIRF predicted, TCN predicted, and measured readout waveforms, with the percent signal difference from the measured value shown for each image. Difference images are scaled to $\pm 60\%$ of the maximum intensity in the measured image. RMSE of images normalized by maximum overall signal are reported.\label{fig5}}
\end{figure}

Figure \ref{fig5} shows magnitude and error images from the phantom acquisitions, reconstructed with all trajectory estimation methods for the rosette and spiral trajectories. Image errors are pixel-wise differences normalized by maximum image intensity. Images reconstructed with the nominal trajectory in both cases exhibited severe artifacts, with dramatic signal intensity dropoff in the periphery of the phantom and hyperintensity in the center. Severe background aliasing was present in both cases. 
The GIRF based reconstruction improved both of these errors, but the TCN based reconstruction resulted in far less signal loss in the phantom.

\begin{figure*}
\centerline{\includegraphics[width=18cm, keepaspectratio]{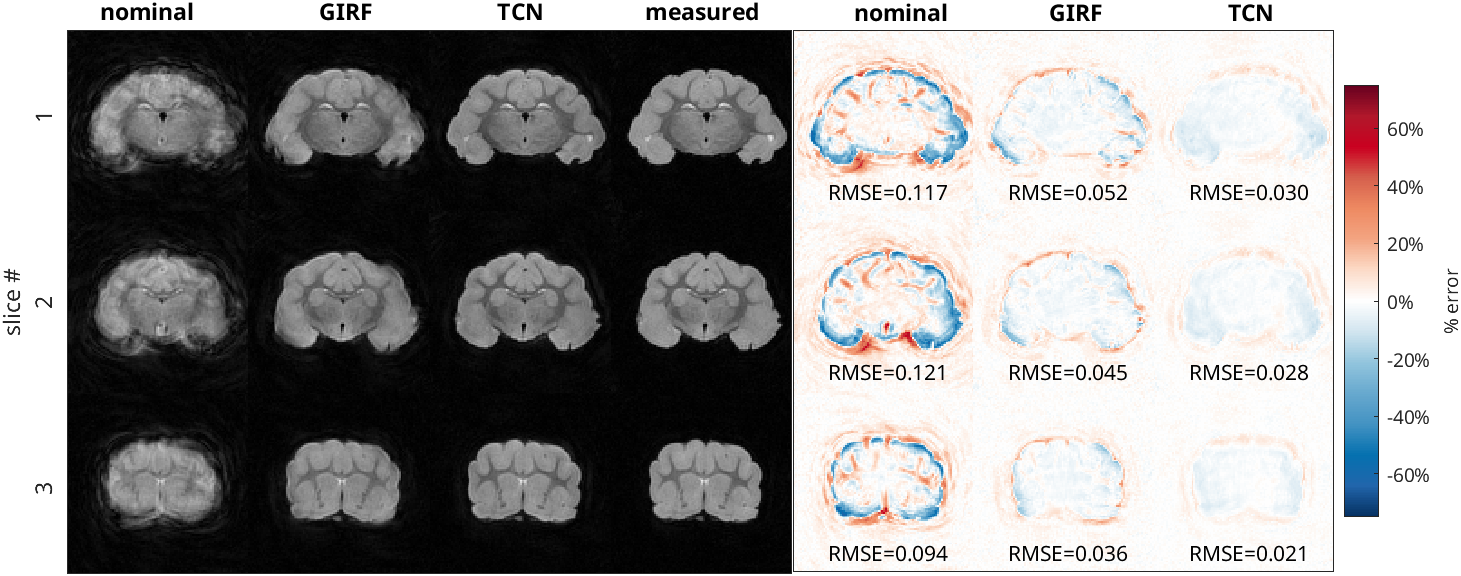}}
\caption{(L)\textit{Ex-vivo} ferret brain magnitude images acquired with a spiral acquisition, for 3 evenly spaced slices. Reconstructions based on the nominal trajectory, GIRF-predicted trajectory, TCN-predicted trajectory, and measured trajectory are shown. (R) Difference images with respect to the measured-trajectory image. Difference images are scaled to $\pm 75\%$ of the maximum intensity in the measured image. RMSE of images normalized by maximum overall signal are reported. \label{fig6}}
\end{figure*}

\begin{figure*}
\centerline{\includegraphics[width=18cm, keepaspectratio]{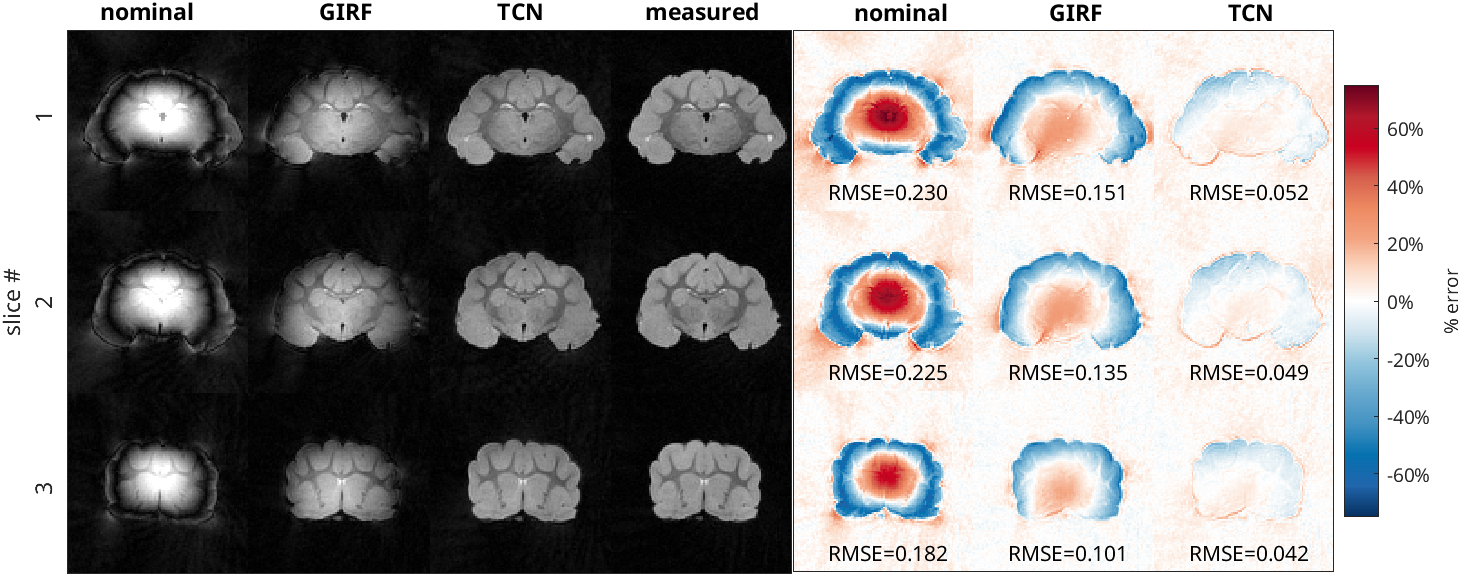}}
\caption{(L) \textit{Ex-vivo} ferret brain magnitude images aquired with a rosette acquisition, for 3 evenly spaced slices. Reconstructions based on the nominal trajectory, GIRF-predicted trajectory, TCN-predicted trajectory, and measured trajectory are shown. (R) Difference images with respect to the measured-trajectory image.  Difference images are scaled to $\pm 75\%$ of the maximum intensity in the measured image. RMSE of images normalized by maximum overall signal are reported.\label{fig7}}
\end{figure*}

The anatomical images of the ferret brain acquired with the spiral trajectory are shown in Figure \ref{fig6}, and the anatomical images of the ferret brain acquired with the rosette trajectory are shown in Figure \ref{fig7}, with three representative slices from each trajectory. As in the phantom images, severe distortions were present in the images reconstructed using the nominal trajectory. The spiral images exhibited severe blurring, while the rosette images primarily had signal intensity artifacts with signal build-up in the center of the image and a ring of signal loss in the periphery. Reconstruction with GIRF-estimated trajectories only partially resolved these effects. Blurring distortions were still evident in the spiral trajectory case, and signal hyperintensities and hypointensities were still present in the rosette trajectory case. Reconstruction with the TCN-predicted trajectories produced images that are much closer to those reconstructed using the measured trajectories, although some artifacts were still present which were not seen in the measured-trajectory images. 

\begin{figure*}
\centerline{\includegraphics[width=18cm, keepaspectratio]{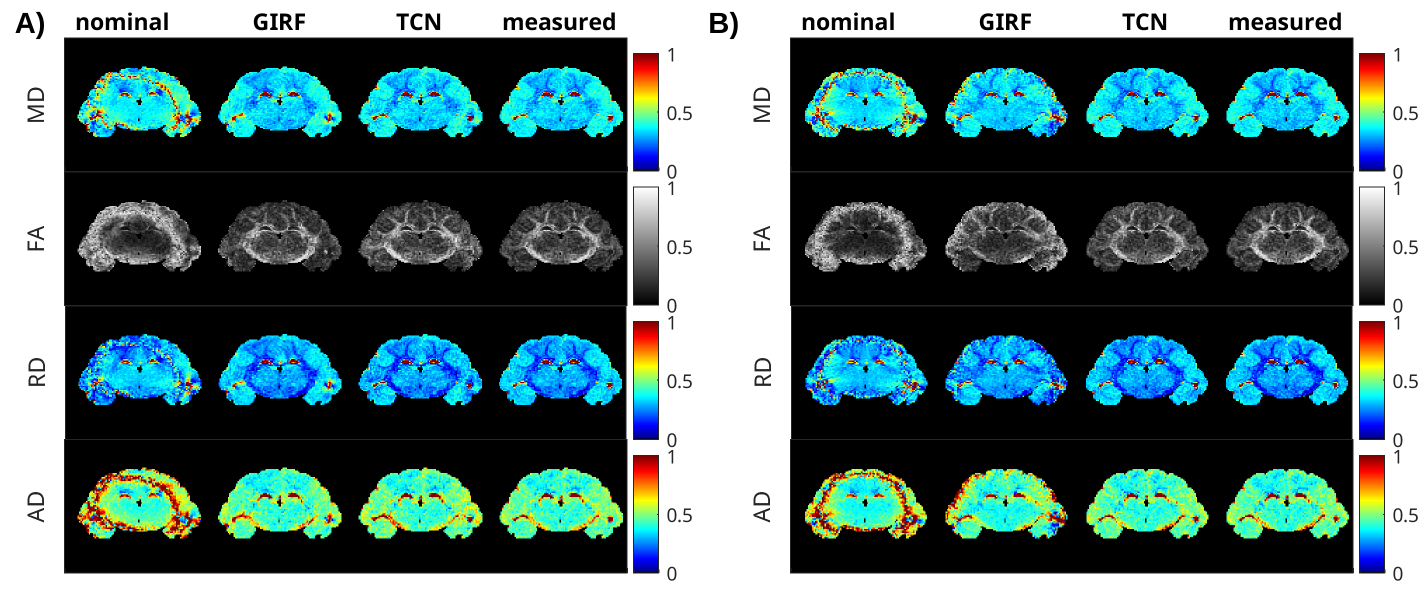}}
\caption{A) Spiral acquired diffusion parameter maps for slice \#1 in Figure \ref{fig6}. B) Rosette acquired diffusion parameter maps for slice \#1 in Figure \ref{fig7}. MD, FA, RD, and AD are shown for images reconstructed using the nominal, GIRF-predicted, TCN-predicted and measured trajectories. \label{fig8}}
\end{figure*}

Diffusion parameter maps in the ferret brain are shown in Figure \ref{fig8}A) and Figure \ref{fig8}B for spiral and rosette acquisitions, respectively. The severe distortions present in the images reconstructed with the nominal trajectories created biased and noisy diffusion parameters in the periphery of the brain in both cases. GIRF and TCN based reconstruction both remediated the most severe errors in parameter estimation. However, diffusion analysis with the GIRF-predicted trajectory images still had a number of significant discrepancies with the parameter maps reconstructed using the measured images. The most notable distortion in the GIRF-reconstructed spiral image was the loss of several white matter tract structures in the FA and RD maps. In the rosette images, the failure of the GIRF to fully resolve signal loss around the periphery of the brain resulted in overestimates of FA and AD and underestimates of RD and MD in those regions. In both cases, reconstruction with TCN-predicted trajectories  produced parameter maps that corresponded more closely to those created using the measured trajectories. 

\section{Discussion}
These results demonstrate the ability of temporal convolutional networks to accurately predict nonlinear imperfections in the gradient system of a small animal MRI system. We show that incorporating these gradient waveform estimates into the reconstruction process provides  downstream improvements in image quality and quantitative parameter estimates. 

We contribute further evidence to the body of work demonstrating that nonlinearities in MRI gradient systems can deviate from linear time-invariant system responses. We document several temporal nonlinearities including zero-crossing distortions, and variable responses to waveform scaling. Such temporal nonlinearities are typically relatively small in clinical systems operating well within their gradient systems' design parameters \cite{Vannesjo2016ImagePrediction}. However, nonlinearities may be more substantial in preclinical systems with high performance gradients \cite{Scholten2024OnInsert, Latta2024Two-parametricMRI}, or systems with relatively low-performance gradient hardware \cite{OReilly2021AMRI, Babaloo2022nonlinearSystems}. Furthermore, even in clinical systems, trajectory errors may be substantial and nonlinear if challenging or nonstandard gradient waveforms are used \cite{Brodsky2009CharacterizingLTI, Harkins2014IterativeWaveforms, Pipe2022GeneratingSpectrum}. Thus, a general model of the gradient system which makes no assumptions about linearity or time invariance is highly desirable.

The GIRF and GSTF provide excellent correction of gradient distortion when temporal nonlinearities are not significant, for example in more conventional Cartesian trajectories which rely on trapezoidal gradient waveforms. Linear distortion correction methods may be sufficient in these settings. The TCN method we introduce broadens the range of correctable gradient distortions, and likely has most additional utility in more challenging imaging readouts or in systems with more limited gradient hardware. We observe that the TCN model performs well in predicting Cartesian trajectory gradient waveforms, including the trapezoidal waveforms in our validation set and the trapezoidal waveforms prepended to the spiral in-out test set waveforms. For Cartesian trajectories, either GIRF or TCN correction may be appropriate.

A lengthy scan session of several hours was used to collect training data for this study, and future work should be aimed at optimizing the library of gradient waveforms and scan time for initial training of such a model. This could include the design of waveforms optimized to provide the necessary characterization of the system in minimal scan time. Some waveforms are likely more important to include than others. For example, a slew rate-constrained chirp (included in our training dataset), is highly desirable in that it covers a broad range of frequencies efficiently. Such a waveform may be more efficient in the improvements it provides  to the model's generalization ability than, for example, a waveform with more limited spectral content such as a multisine (also included in our training data). 

\begin{figure*}
\centerline{\includegraphics[width=16cm, keepaspectratio]{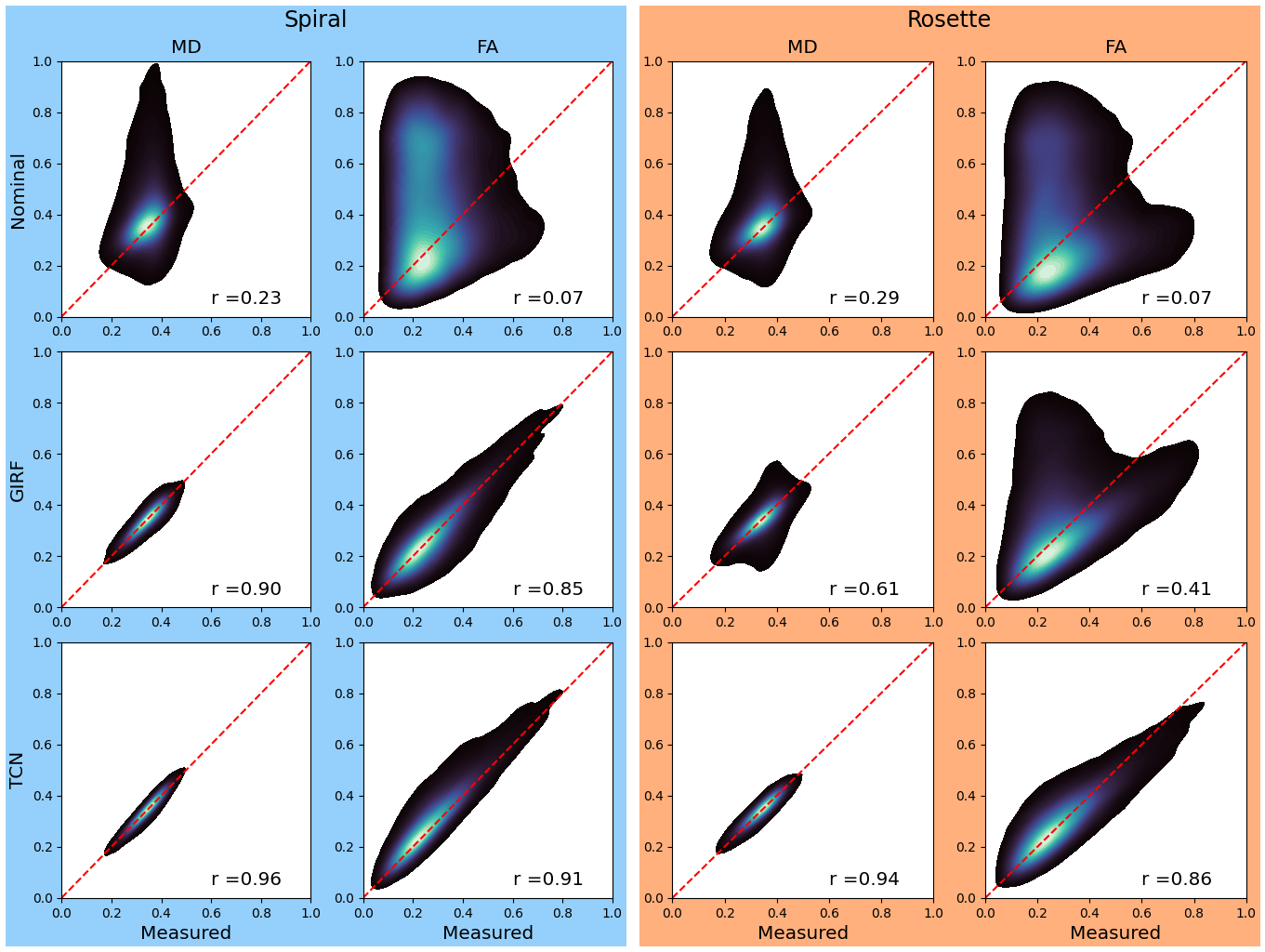}}
\caption{Diffusion parameter correlation between the measured trajectory ('ground truth') values and values derived from reconstruction with various trajectory prediction methods. Individual plots are a kernel density estimation of pixel-wise parameter values thresholded at 0.05, with correlation coefficient reported.\label{fig10}}
\end{figure*}

The model's ability to predict the gradient system's response to previously unseen gradient waveforms has the potential to eliminate the need for application-specific scan-time calibration sequences. Further, once a model is initially trained, a subset of waveforms can likely be used to update the model periodically. Previous work has shown that the responses of the gradient system remain relatively stable over time \cite{Vannesjo2016ImagePrediction}.

The TCN architecture was selected due to its excellence in capturing long-term temporal relationships in time series data\cite{Lea2017TemporalDetection}, and its relative ease of training compared to recurrent neural networks \cite{Pascanu2013OnNetworks, Bai2018AnModeling}. Further optimization of the network architecture and comparison to other architectures appropriate for sequence modeling is possible. 

Several limitations to this study should be noted. First, we created an independent model for each gradient axis. As such, the model does not incorporate the possibility of correcting for cross-term responses. The training data acquisition process could be expanded to include measurements made while multiple gradient chains are active, although this would further lengthen the training data acquisition stage. Furthermore, since the TCN model is based on measurements made on a static phantom, it is not able to predict any gradient response changes that may be subject-specific or dynamic across a scan. For applications in which these are significant considerations, concurrent monitoring such as that provided by a field camera may be preferred \cite{Dietrich2016AAnalysis}.

\par It is also known that gradient hardware heating, common in long scans, may result in temperature-dependent changes in the gradient system response \cite{Nussbaum2022ThermalModeling, Stich2020TheCharacteristics}. In order to model non-LTI changes in the gradient response over a scanning session, such as those resulting from hardware temperature changes, modifications to the TCN network would be required. For example, an input feature related to the gradient duty cycle or measured gradient hardware temperature could be included. The training data for these experiments was acquired in one long, uninterupted session, so it is likely that our trained model actually reflects the performance of the gradient system at high temperatures as most data was likely acquired after the scanner hardware reached a high-temperature. The imaging experiments were performed in a separate imaging session from an initial "cold" hardware state, so our TCN model likely includes error due to unaccounted for temperature variations. Future investigations could attempt to include hardware temperature as a feature to more accurately model time-varying hardware performance.

In the imaging experiments, gradient waveform prediction was  applied only to the readout waveforms. Gradient waveforms used for non-cartesian readouts are particularly demanding due to their relatively long duration and time-varying envelopes, which make them a natural subject of this initial study. However, it is possible that very large gradient system imperfections could also impact other gradient waveforms, such as diffusion-encoding gradients, phase-encoding lobes, or slice-select gradients. Since these were all undemanding trapezoidal waveforms in this study, distortions were assumed to be minimal relative to those manifesting in the more challenging non-Cartesian readouts. However, temporal gradient distortions can also affect diffusion encoding\cite{Szczepankiewicz2019Maxwell-compensatedEncoding}, and excitation localization \cite{Harkins2014IterativeWaveforms}, particularly when more demanding gradient waveforms are used. Further studies could advance the work done here on improving diffusion parameter mapping by applying a prediction model to the diffusion encoding waveforms themselves.
Incorporating the effects of diffusion gradients on the imaging readout would likely require a separate model trained on waveform measurements made following diffusion gradients, to characterize long-time constant eddy current effects. More complex diffusion gradient waveforms than the PGSE gradients used in this work, such as oscillating gradient spin echo (OGSE) encoding gradients, could themselves be subject to severe distortions due to nonideal gradient hardware. Deviation from nominal OGSE diffusion waveforms could affect the diffusion parameters being measured. Future studies could investigate estimation of diffusion parameters using nonideal OGSE waveforms, but that exceeds the scope of this initial work.

We emphasize that as described, the proposed method accounts only for temporal nonlinearities in the gradient waveforms. Accounting for spatially nonlinear and cross terms would involve both changes to the gradient waveform measurement sequence and modifications of the TCN model, which we consider outside the scope of the current report. Some prior studies have investigated machine learning techniques to spatial gradient nonlinearity correction \cite{Shan2023DistortioncorrectedMRILinac, Hu2020DistortionDeep-learning}, and hybrid approaches that account for both distortion types could be investigated.

There are general limitations to a retrospective prediction approach to gradient errors as opposed to prospective methods meant to preemptively prevent errors in the gradient waveforms. For example, errors in the excitation profile produced by trajectory errors cannot usually be corrected retrospectively. Our model of the gradient system response is highly accurate, so the model could be used within iterative predistortion approaches for offline design of preemphasized waveforms \cite{Harkins2014IterativeWaveforms, Fillmer2016FastFMRI}. Machine learning-based approaches are also being examined to determine MRI waveform pre-emphasis for situations in which this retrospective approach is not appropriate \cite{Albert2024RFModel, Martin2024GradientLearning}. 

\section{Conclusion}

In conclusion, a framework for using deep learning has been presented to remove the impact of readout gradient waveform distortions from image reconstruction and quantiative diffusion parameter estimation. This method may enable the use of more advanced and challenging gradient waveforms in MRI that would otherwise be prohibited by system imperfections.

\section*{Acknowledgments}
The authors gratefully acknowledge Daniel Colvin for support with the Bruker imaging systems. This study was funded by the following grants from the National Institute of Health: R01 EB031954 and T32 EB003817. 

\subsection*{Financial disclosure}

None reported.

\subsection*{Conflict of interest}

The authors declare no potential conflict of interests.

\bibliography{MRM-AMA}%
\vfill\pagebreak




\vspace*{6pt}

\end{document}